\documentclass[12pt]{article}

\usepackage{multirow} 
\usepackage{fullpage}
\usepackage{latexsym}              
\usepackage{amssymb}               
\usepackage{amsmath}               
\usepackage{amsthm}                
\usepackage{bbm}
\usepackage{url}
\usepackage{rotating}
\usepackage{caption}
\usepackage{setspace}
\usepackage{thmtools}

\usepackage{arydshln}

\usepackage{subcaption}
\usepackage{graphicx}
\usepackage[english]{babel}

\newcommand{\blind}{0}
\newtheorem{theorem}{Theorem}[section]
\newtheorem{lemma}[theorem]{Lemma}

\declaretheoremstyle[]{normalhead}

\newcommand{\bb}{{\bf b}}

\newcommand{\by}{{\bf y}}

\newcommand{\bw}{{\bf w}}
\newcommand{\bs}{{\bf s}}
\newcommand{\bt}{{\bf t}}

\newcommand{\bA}{{\bf A}}

\newcommand{\bC}{{\bf C}}

\newcommand{\bI}{{\bf I}}
\newcommand{\bK}{{\bf K}}

\newcommand{\bR}{{\bf R}}
\newcommand{\bS}{{\bf S}}
\newcommand{\bU}{{\bf U}}
\newcommand{\bV}{{\bf V}}
\newcommand{\bW}{{\bf W}}
\newcommand{\bX}{{\bf X}}

\newcommand{\bZ}{{\bf Z}}

\newcommand{\bone}{{\bf 1}}
\newcommand{\bzero}{{\bf 0}}
\newcommand{\bo}{\boldsymbol}

\newcommand{\bSigma}{{\boldsymbol\Sigma}}

\newcommand{\bbeta}{{\boldsymbol\beta}}

\newcommand{\beq}{\begin{equation}}
\newcommand{\eeq}{\end{equation}}
\newcommand{\inv}{^{-1}}
\newcommand{\bXp}{{\bf X}_{pri}}
\newcommand{\bXq}{{\bf X}_{pot}}
\newcommand{\betap}{{\bo\beta}_{pri}}
\newcommand{\betaq}{{\bo\beta}_{pot}}

\makeatletter

\newcount\mn \newcount\hr
\def\timeofday{%
  \hr=\time \divide\hr 60
  \mn=-\hr \multiply\mn 60 \advance\mn \time
  \ifnum\hr=0%
     {12\,:\,\twodigits\mn\,am}%
  \else{%
     \ifnum\hr<12%
        {\number\hr\,:\,\twodigits\mn\,am}%
     \else{%
        \ifnum\hr=12%
          {\number\hr\,:\,\twodigits\mn\,pm}%
        \else%
          {\advance\hr -12 \number\hr\,:\,\twodigits\mn\,pm}%
        \fi}%
     \fi}%
  \fi}
\def\twodigits#1{\ifnum #1<10 0\fi \number#1}
\def\tersetoday{\ifcase\month\or
  Jan.\or Feb.\or Mar.\or Apr.\or May\or June\or
  July\or Aug.\or Sep.\or Oct.\or Nov.\or Dec.\fi
  \space\number\day, \number\year}

\date{}

\begin{document}
  \title{\bf Generalized Bayesian $D$ criterion for single-stratum and multistratum designs}

\if0\blind
{
   \author{Chang-Yun Lin\\    \small \it Department of Applied Mathematics and Institute of Statistics,\\ \small \it National Chung Hsing University, Taichung, Taiwan, 40227     
 }\fi
    
\maketitle

\begin{abstract}
DuMouchel and Jones (1994) proposed the Bayesian $D$ criterion by modifying the $D$-optimality approach to reduce dependence of the selected design on an assumed model. 
This criterion has been applied to select various single-stratum designs for completely randomized experiments. 
In many industrial experiments, complete randomization is sometimes expensive or infeasible and, hence, designs used for the experiments often have multistratum structures. 
However, the original Bayesian $D$ criterion was developed under the framework of single-stratum structures and cannot be applied to select multistratum designs. 
In this paper, we study how to extend the Bayesian approach for more complicated experiments and develop the generalized Bayesian $D$ criterion, which generalizes the original Bayesian $D$ criterion and can be applied to select single-stratum and multistratum designs for various experiments. 
 \end{abstract}


\section{Introduction}
The $D$ criterion is commonly used in design of experiments to select optimal designs. 
The usual statistical justification for the $D$-optimality is that it minimizes the volume of the joint confidence region for the coefficients in a model (Atkinson et al., 2007, p. 135). 
%
However, $D$-optimal designs have been criticized for being too dependent on an assumed model. 
To reduce the dependence, DuMouchel and Jones (1994) proposed a simple Bayesian modification for the $D$-optimality approach. 
They assumed that, in addition to the primary terms that are fitted, there exist potential terms that are possibly important. 
Typically, the sample size is not large enough to estimate all of the two terms simultaneously. 
The Bayesian approach adds prior information to avoid a singular estimation problem and, hence, allows the precise estimation for all the primary terms while providing detectability for the potential terms. 
This approach has been applied to construct and select optimal designs for various completely randomized experiments.  
For instance, Andere-Rendon et al. (1997) used it to select designs for mixture experiments,  Lin et al. (2000) applied it to construct Bayesian two-stage optimal designs for mixture models, Ruggoo and Vandebroek (2004, 2006) used it to generate sequential model-robust designs, Jones et al. (2008) applied it to construct Bayesian $D$-optimal supersaturated designs, and Gutman et al. (2014) developed an augmenting method with the Bayesian $D$ criterion by adding runs to existing supersaturated designs.

The Bayesian $D$-optimal designs mentioned above were basically developed for completely randomized experiments. They have single-stratum structures and can be only used for the experiments whose run orders are completely randomized.  
However, in many real cases, experiments are often more complicated so that sometimes the complete randomization is expensive or infeasible. 
%
For instance, many industrial experiments contain factors whose levels are hard to change. If complete randomization is conducted for such experiments, the cost will increase due to frequently changing levels of these factors. To reduce cost, experimenters often conduct a randomization for level combinations of the hard-to-change factors and then implement another randomization for level combinations of the easy-to-change factors. This two-stage randomization forms a two-stratum structure and the designs developed for such experiments are called the split-plot designs (see Section~\ref{se:sp}). 
Another multistratum structure often seen in industrial experiments is the strip-plot structure. 
If an experiment contains two distinct stages where the groups of semi-products produced in the first stage are applied to the second stage, then the two-stage process results in a three-stratum structure. 
Designs developed for such experiments are called the strip-plot designs (see Section~\ref{se:st}).

However, the Bayesian $D$ criterion proposed by DuMouchel and Jones (1994) was developed under the framework of single-stratum structures and cannot be applied to select designs which have multistratum structures. 
This motivates us to extend the Bayesian approach for more complicated experiments. 
We consider more general cases and develop a generalized Bayesian $D$ criterion which can be used for selecting optimal designs for single-stratum or multistratum experiments. 
%
The remainder of this article is organized as follows. Section~\ref{se:method} introduces the model for multistratum designs and develops the generalized Bayesian $D$ criterion. Section~\ref{se:construction} provides a scaling method and a briefly algorithm for constructing and searching for generalized Bayesian $D$-optimal designs. 
The applications, comparisons, and sensitivity analyses of the generalized Bayesian $D$ criterion for split-plot, strip-plot, and staggered-level designs are provided in Section~\ref{se:app}. 
Section~\ref{se:cr} is the concluding remarks.

\section{The method and criterion}\label{se:method}
\subsection{The model and the estimation problem}\label{se:model}
Consider an $n\times m$ design in a $g$-stratum structure and assume that there are $b_l$ units and $m_l$ factors in the $l$th stratum, where $l=1,\cdots,g$, $n=b_g$ and $m=\sum_{l=1}^gm_l$. 
In general, there are $p$ primary terms that we want to fit and the model under the $g$-stratum structure can be expressed as 
\begin{equation}\label{eq:mod}
\by=\bX_{pri}\bo\beta_{pri}+\sum_{l=1}^g\bU_l\bo\gamma_l,
\end{equation} 
where $\by$ is an $n\times 1$ vector of the responses, $\bo\beta_{pri}$ is the $p\times 1$ vector of effects for the primary terms, $\bX_{pri}$ is the $n\times p$ model matrix for $\bo\beta_{pri}$, $\bo\gamma_l$ is the $b_l\times 1$ vector of random effects for the $l$th stratum, and $\bU_l$ is the $n\times b_l$ indicator matrix for $\bo\gamma_l$. 
Note that $\bU_g=\bI_n$, where $\bI_n$ is an $n\times n$ identity matrix. Hence, for single-stratum structure, i.e., $g=1$, model (\ref{eq:mod}) can be written as $\by=\bX_{pri}\bo\beta_{pri}+\bo\epsilon$, where $\bo\epsilon=\bo\gamma_1$, and for $g\geq 2$, it can be written as $\by=\bX_{pri}\bo\beta_{pri}+\sum_{l=1}^{g-1}\bU_l\bo\gamma_l+\bo\epsilon$, where $\bo\epsilon=\bo\gamma_g$. 
Assume that $\bo\gamma_l\sim N(\bzero_{b_l}, \sigma_l^2\bI_{b_l})$ and that $\bo\gamma_i$ and $\bo\gamma_j$ are independent for $i\neq j$, where $\bzero_{b_l}$ is a $b_l\times 1$ vector of zeros. Let $\eta_l=\sigma_l^2/\sigma_g^2$ for $l=1,\cdots,g$. Then the information matrix for $\bo\beta_{pri}$ is $\bX_{pri}'\bo\Sigma\inv\bXp$, where $\bo\Sigma=\sigma_g^2\sum_{l=1}^g\eta_l\bU_l\bU_l'$ is the covariance matrix of $\by$. 
The generalized least squares (GLS) estimator for $\betap$ is $(\bXp'\bo\Sigma\inv\bXp)\inv\bXp'\bSigma\inv\by$ and the covariance matrix of the GLS estimator is $(\bXp'\bo\Sigma\inv\bXp)\inv$. 
The $D$ criterion selects designs which maximize $|\bX_{pri}'\bo\Sigma\inv\bXp|^{1/p}$, the $p$th root of the determinant of the information matrix for $\betap$.  

Consider now that, in addition to the $p$ primary terms, there are $q$ potential terms that are possibly important.  
Let $\betaq$ be the $q\times 1$ vector of effects for the potential terms and $\bXq$ be the $n\times q$ model matrix for $\betaq$. Then the model including both primary and potential terms under the $g$-stratum structure can be expressed by 
\begin{equation}\label{eq:mod1}
\by=\bX\bo\beta+\sum_{l=1}^g\bU_l\bo\gamma_l,
\end{equation} 
where $\bX=(\bXp,\bXq)$ and $\bo\beta=(\betap',\betaq')'$.  
In many cases, the rank of $\bX$, denoted by $rank(\bX)$, is usually not large enough for estimating all the $p+q$ effects simultaneously because, typically, $p<rank(\bX)<p+q$ and, hence, $\bX'\bo\Sigma\inv\bX$ is singular. In this situation, the determinant of the information matrix for $\bo\beta$ is zero.  
Therefore, it is necessary to add prior information to circumvent the singular estimation problem.

\subsection{The Bayesian approach and criterion}
Following the Bayesian assumption in DuMouchel and Jones (1994), we add prior information for $\bo\beta$ in  model (\ref{eq:mod1}).  
Since the primary terms are likely to be active, the effects corresponding to the primary terms are assumed to have a prior mean of zero and a prior variance tending to infinity, i.e., $\betap\sim N(\bzero_p,\xi^2\bI_p)$ where $\xi^2\rightarrow \infty$. 
On the other hand, the potential terms are unlikely to have large effects. Hence, the effects corresponding to the potential terms are assumed to have a prior mean of zero and a finite variance, i.e., $\betaq\sim N(\bzero_q,\tau^2\bI_q)$. Assume that $\betap$, $\betaq$, and $\bo\gamma_l$ for $l=1,\cdots,g$ are mutually independent. Combing $\betap$ and $\betaq$, we rewrite the above priors as $\bo\beta\sim N(\bzero_{r}, \bR)$ where $r=p+q$ and 
\begin{equation}
\bR=\left(\begin{array}{cc}
\xi^2\bI_{p}&\bzero_{p\times q}\\
\bzero_{q\times p}&\tau^2\bI_{q}
\end{array}\right).
\end{equation}
%
Since $\bo\gamma_l$ and $\bo\beta$ are uncorrelated, we obtain $cov(\bbeta,\by)=cov(\bo\beta,\bX\bo\beta)=\bR\bX'$ and $cov(\by)=cov(\bX\bo\beta,\bX\bo\beta)+\sum_{l=1}^gcov(\bU_l\bo\gamma_l,\bU_l\bo\gamma_l)=\bX\bR\bX'+\bSigma$. Also, $E(\bbeta)=\bzero_r$, $cov(\bbeta)=\bR$, and $E(\by)=\bzero_n$. 
Therefore, the joint probability of $\bbeta$ and $\by$ is given by 
\begin{equation}
\left[\begin{array}{c}
\bbeta\\
\by
\end{array}\right]\sim N\left(\left[\begin{array}{c}
\bzero_r\\
\bzero_n
\end{array}\right],\left[\begin{array}{cc}
\bR&\bR\bX'\\
\bX\bR&\bX\bR\bX'+\bSigma
\end{array}\right]
\right).
\end{equation}
By Lemma B.1.2 in Santner et al. (2003, p. 211), 
we obtain the posterior distribution $\bbeta |\by\sim N(\bb,\bS)$, where
\begin{equation}\label{eq:mu_b|y}
\bb=\bR\bX'(\bX\bR\bX'+\bSigma)\inv\by
\end{equation}
and
\begin{equation}\label{eq:sig_b|y}
\bS=\bR-\bR\bX'(\bX\bR\bX'+\bSigma)\inv\bX\bR
\end{equation}
are the Bayesian estimator and posterior variance of $\bbeta$, respectively. 
Since $\xi^2$ approaches infinity, we have $\bR\inv=\bK/\tau^2$, where 
\[
\bK=\left(\begin{array}{cc}
\bzero_{p\times p}&\bzero_{p\times q}\\
\bzero_{q\times p}&\bI_{q}\\
\end{array}\right).
\]
By the following two equalities: (i) $\bA\inv\bU(\bV\bA\inv\bU+\bC\inv)\inv=(\bU\bC\bV+\bA)\inv\bU\bC$ and (ii) $\bA\inv-\bA\inv \bU(\bV\bA\inv \bU+\bC\inv)\inv \bV\bA\inv=(\bU\bC\bV+\bA)\inv$ (Woodbury's formula, see Harville 1997, p. 424), Equations (\ref{eq:mu_b|y}) and (\ref{eq:sig_b|y}) can be further simplified as 
\begin{equation}\label{eq:mu_b|y1}
\bb=(\bX'\bSigma\inv\bX+\bK/\tau^2)\inv\bX'\bSigma\inv\by
\end{equation}
and 
\begin{equation}\label{eq:sig_b|y1}
\bS=(\bX'\bSigma\inv\bX+\bK/\tau^2)\inv.
\end{equation}
For single-stratum designs ($g=1$), the covariance matrix of $\by$ reduces to $\bSigma=\sigma_1^2\bI_n$. 
When $\sigma_1^2=1$ as set in DuMouchel and Jones (1994), Equations (\ref{eq:mu_b|y1}) and (\ref{eq:sig_b|y1}) reduce to $\bb=(\bX'\bX+\bK/\tau^2)\inv\bX'\by$ and $\bS=(\bX'\bX+\bK/\tau^2)\inv$, respectively, which are identical to the Bayesian results given in DuMouchel and Jones (1994) (see also Box and Tiao, 1973, and Lee, 1989) for single-stratum designs.

We define the criterion which selects designs with minimum $|\bX'\bSigma\inv\bX+\bK/\tau^2|^{1/r}$ as the generalized Bayesian $D$ (GBD) criterion. 
When $g=1$, the GBD criterion minimizes $|\bX'\bX+\bK/\tau^2|^{1/r}$ and, hence, is equivalent to the Bayesian $D$ criterion proposed by DuMouchel and Jones (1994). 
The GBD criterion is easy to use and has the following advantages. 
First, it reduces dependence of $D$-optimal designs on an assumed model by considering the primary and potential terms simultaneously. The GBD-optimal designs allow the precise estimation for all the primary terms while providing detectability for the potential terms.  
Second, by adding $\bK/\tau^2$, the GBD criterion avoids the singularity problem for $\bX'\bSigma\inv\bX$ when rank of $\bX$ is smaller than $p+q$. 
Third, it generalizes the Bayesian $D$ criterion proposed by DuMouchel and Jones (1994) and, hence, can be applied to determine optimal single-stratum and  multistratum designs for various experiments. 




\section{Algorithm and construction method}\label{se:construction}
\subsection{Preliminary scaling and centering}
To eliminate the aliasing over the candidates and permit the use of the prior distribution to work well, we apply the scaling and centering method proposed in DuMouchel and Jones (1994) as follows:  
(i) each primary term (except the intercept) ranges from $-1$ and $+1$, i.e., $min\{\bXp\}=-1$ and $max\{\bXp\}=1$, (ii) the range of each potential term is unity, i.e., $max\{\bXq\}-min\{\bXq\}=1$, and (iii) each potential term is approximately uncorrelated with all primary terms, i.e., $\sum\bXp\bXq=0$, where the summation is taken over the set of candidate points. 
To achieve this scaling, we compute $\bW=\bXq-\bXp\bo\alpha$, where $\bo\alpha=(\bXp'\bXp)\inv\bXp'\bXq$, and let $\bZ=\bW/(max\{\bW\}-min\{\bW\})$. Then replace the potential term $\bXq$ by $\bZ$. 
The reader is referred to DuMouchel and Jones (1994) for more details.  

\subsection{Algorithm}
To obtain the GBD-optimal design, we apply the coordinate-exchange algorithm proposed by Jones and Goos (2007). 
Let $d$ denote the value of $|\bX'\bSigma\inv\bX+\bK/\tau^2|^{1/r}$ of a design, where $\bXq$ in $\bX$ is replaced by $\bZ$ according to the scaling and centering method. Let $D_{cur}$ and $D_{opt}$ denote the current and optimal design, respectively, during the search, $d_{cur}$ and $d_{opt}$ denote the $d$ values of the current and optimal designs, respectively, $t_{total}$ be the total number of random starts we want to implement (usually, $t_{total}>10^5$ produces better results), $t$ be the $t$th random start, and $q=1$ or $0$ indicate whether the design is improved or not after the coordinate exchange.    
Since the algorithm is a simple modification of that given in Jones and Goos (2007), we briefly describe it as follows. 
\begin{itemize}
\item[ ] \begin{itemize}
\item[Step 1.] Set $t=1$ and $d_{opt}=0$.  
\item[Step 2.] Set $q=0$. Randomly generate a starting design, $D_0$, according to the multistratum structure and compute its $d$ value, $d_0$. Set $D_{cur}=D_0$ and $d_{cur}=d_0$. 
\item[Step 3.] Sequentially implement coordinate exchange for $D_{cur}$ from stratum $l=1$ to $l=g$ to improve the current design. For each updated design $D^*$ obtained by the coordinate exchange, calculate its $d$ value, $d^*$. If $d^*>d_{cur}$, then let $q=1$ and set $D_{cur}=D^*$ and $d_{cur}=d^*$. 
\item[Step 4.] If $q=1$, set $q=0$ and go back to Step 3. Otherwise, go to Step 5. 
\item[Step 5.] If $d_{cur}>d_{opt}$, set $D_{opt}=D_{cur}$ and $d_{opt}=d_{cur}$. Go to Step 6.
\item[Step 6.] If $t<t_{total}$, let $t=t+1$ and go back to Step 2. Otherwise, end the search. 
\end{itemize}
\end{itemize}
%
The final $D_{opt}$ obtained through this procedure is reported as the GBD-optimal design. 

\subsection{Choice of $\tau$}\label{se:tau}
The Bayesian approach assumes that the effects of potential terms have a prior variance $\tau^2$. 
When $\tau$ approaches zero, which implies that the effects from the potential terms are very small or do not even exist, the resulting design obtained by the GBD criterion is equivalent to the optimal design obtained by the $D$ criterion for the model including primary terms only.   
%
If experimenters suspect that the effects of potential terms are more likely to be significant, a larger $\tau$ should be assigned. 
Since the variance of the response is $\sigma_y^2=\sigma_g^2\sum_{l=1}^g\eta_l$, it is reasonable to choose  $\tau\geq 3\sigma_y$.  
Note that $\sigma_g^2$ does not affect the result of the search and, hence, can be set as one without loss of generality.   
%
If $\eta_l$'s are unknown, setting $\eta_l=10$ for $l=1,\cdots,g-1$ (note that $\eta_g=\sigma_g^2/\sigma_g^2=1$) can be considered. 
This is because $\eta_l$ is less than or equal to 10 for $l=1,\cdots,g-1$ in most cases presented in the literature and setting $\eta_l$'s at maximum values permits the resulting design to perform well on detecting active potential terms. 
Hence, it is suggested to choose $\tau=10$ ($\approx 3\sqrt{max(\eta_1)+\eta_2}=3\sqrt{10+1}$) for two-stratum designs and $\tau=14$ ($\approx 3\sqrt{max(\eta_1)+max(\eta_2)+\eta_3}=3\sqrt{10+10+1}$) for three-stratum designs.   

    


\section{Applications, comparisons, and sensitivity analyses}\label{se:app}

In this section, we provide three examples to show the advantages of the GBD criterion on constructing multistratum designs. 
Section~\ref{se:sp} illustrates how to apply the GBD criterion on selecting multistratum designs, which cannot be accomplished by the original Bayesian $D$ criterion. 
Section~\ref{se:st} compares $D$- and GBD-optimal multistratum designs and shows that GBD-optimal designs provide better detectability for the potential terms than $D$-optimal designs.    
Section~\ref{se:sl} demonstrates how to use the GBD criterion to construct multistratum designs whose rank are not large enough for estimating all of the coefficients in the model, which cannot be achieved by using the $D$ criterion.     
Sensitivity analyses are provided in the end of each section.

\subsection{Split-plot designs}\label{se:sp}
Split-plot designs are often used when experiments contain whole-plot factors, whose levels are hard to change, and sub-plot factors, whose levels are easy to change. 
The level combinations of the whole-plot and subplot factors are called the whole plots and subplots, respectively.  
A two-step randomization implemented on the whole plots and subplots separately makes a split-plot design form a two-stratum structure. Such randomization reduces the frequency of level change for the whole-plot factors and, hence, can efficiently save experimental cost. The reader is referred to Goos and Vandebroek (2001, 2003, 2004) and Jones and Nachtsheim (2009) for details. 
%

DuMouchel and Jones (1994) proposed the Bayesian $D$ criterion and applied it to select single-stratum designs for completely randomized experiments.    
In Example~3 of that paper, the authors considered four quantitative factors, A, B, C, and D, with sample size $n=9$ and supposed that the $p=5$ primary terms constituted the first-order model.  
They took a $3^4$-candidate set $=\{-1,0,1\}^4$ to generate optimal designs under the following scenarios: (i) no potential terms (using the $D$ criterion), (ii) square of the factors as the potential terms, (iii) two-factor interactions as the potential terms, (iv) both square and two-factor interaction terms as the potential terms. 
The optimal single-stratum designs under the four scenarios are given in Table~1 of DuMouchel and Jones (1994). 
Since the Bayesian $D$ criterion is a special case of the GBD criterion, this work can be also accomplished by using the GBD criterion with $g=1$ and $\bSigma=\bI_n$.
%
\begin{table}
\begin{center}
\caption{Four optimal split-plot designs for the four scenarios in Section~\ref{se:sp}.}\label{tb:sp9}

\begin{tabular}{ccccc}							
\begin{tabular}{c}							
\\							
WP\\							
1\\							
1\\							
1\\							
2\\							
2\\							
2\\							
3\\							
3\\							
3\\							
\end{tabular}							
&							
\begin{tabular}{r|rrr}							
\multicolumn{4}{c}{$D_{sp1}$}\\							
\hline							
A	&	B	&	C	&	D	\\
\hline							
1	&	1	&	1	&	1	\\
1	&	1	&	-1	&	-1	\\
1	&	-1	&	-1	&	-1	\\
\hline							
1	&	1	&	1	&	-1	\\
1	&	-1	&	1	&	1	\\
1	&	-1	&	-1	&	1	\\
\hline							
-1	&	1	&	-1	&	1	\\
-1	&	1	&	-1	&	-1	\\
-1	&	-1	&	1	&	-1	\\
\hline							
\end{tabular}							
&							
\begin{tabular}{r|rrr}							
\multicolumn{4}{c}{$D_{sp2}$}\\							
\hline							
A	&	B	&	C	&	D	\\
\hline							
1	&	1	&	1	&	1	\\
1	&	0	&	0	&	-1	\\
1	&	-1	&	-1	&	0	\\
\hline							
0	&	1	&	-1	&	-1	\\
0	&	0	&	1	&	0	\\
0	&	-1	&	0	&	1	\\
\hline							
-1	&	1	&	0	&	0	\\
-1	&	0	&	-1	&	1	\\
-1	&	-1	&	1	&	-1	\\
\hline							
\end{tabular}							
&							
\begin{tabular}{r|rrr}							
\multicolumn{4}{c}{$D_{sp3}$}\\							
\hline							
A	&	B	&	C	&	D	\\
\hline							
1	&	1	&	1	&	1	\\
1	&	1	&	-1	&	-1	\\
1	&	-1	&	1	&	-1	\\
\hline							
1	&	1	&	1	&	-1	\\
1	&	-1	&	-1	&	1	\\
1	&	-1	&	-1	&	-1	\\
\hline							
-1	&	1	&	-1	&	1	\\
-1	&	-1	&	1	&	1	\\
-1	&	-1	&	-1	&	-1	\\
\hline							
\end{tabular}							
&							
\begin{tabular}{r|rrr}							
\multicolumn{4}{c}{$D_{sp4}$}\\							
\hline							
A	&	B	&	C	&	D	\\
\hline							
1	&	1	&	1	&	-1	\\
1	&	1	&	-1	&	1	\\
1	&	-1	&	1	&	1	\\
\hline							
0	&	1	&	1	&	1	\\
0	&	0	&	0	&	0	\\
0	&	-1	&	-1	&	-1	\\
\hline							
-1	&	1	&	-1	&	-1	\\
-1	&	-1	&	1	&	-1	\\
-1	&	-1	&	-1	&	1	\\
\hline							
\end{tabular}							
\end{tabular}							
\end{center}
\end{table}

We now consider a more complicated situation in which factor A is a whole-plot factor. 
To reduce the frequency of the level change for factor A, the experimenters want to use a split-plot design with three whole plots and three subplots in each whole plot. 
Assume that $\eta_1=1$. 
To generate the optimal split-plot designs for experiments under the four scenarios, we apply the $D$ criterion for scenario (i) and the GBD criterion with $\tau=10$ for the others. 
The candidate set is $\{-1,0,1\}^4$ and the resulting designs are listed in Table~\ref{tb:sp9}. 
The first design, $D_{sp1}$, in Table~\ref{tb:sp9} is a $D$-optimal split-plot design for the first-order model. It reduces to a two-level design and has no factor settings at 0. 
The second design, $D_{sp2}$, in Table~\ref{tb:sp9} shows the GBD-optimal split-plot design when the $q=4$ square terms are the potential terms. This design is a Latin square (L9) design and identical to the second design in Table~1 of DuMouchel and Jones (1994). 
In this case the model matrix $\bX=(\bX_{pri},\bX_{pot})$ has full rank ($n=p+q$) so that the $D$ criterion for the nine-term model can be calculated, resulting in the same design. 
The third design, $D_{sp3}$, in Table~\ref{tb:sp9} is the GBD-optimal split-plot design when the potential terms are chosen as the $q=6$ two-factor interactions (without square terms). 
Similar to the $D$-optimal designs for scenario (i), $D_{sp3}$ has no factor settings at 0 and reduces to a two-level design. 
The last design, $D_{sp4}$, in Table~\ref{tb:sp9} is the GBD-optimal split-plot design when both square terms and interactions are the potential terms. 
This design contains a center run and four pairs of foldover runs. 
%

\begin{table}
\begin{center}
\caption{Efficiency comparison for the four split-plot designs in Table~\ref{tb:sp9}.}\label{tb:sp.eff}
\begin{tabular}{cccccc}											
\hline											
Scenario	&	Pot. terms	&	$D_{sp1}$	&	$D_{sp2}$	&	$D_{sp3}$	&	$D_{sp4}$	\\
\hline											
(i)	&	No pot.	&	1	&	.785	&	.985	&	.881	\\
(ii)	&	Squares	&	.126	&	1	&	.125	&	.328	\\
(iii)	&	Interactions	&	.972	&	.447	&	1	&	.759	\\
(iv)	&	Both	&	.888	&	.884	&	.906	&	1	\\
\hline											
\end{tabular}																						
\end{center}
\end{table}

Table~\ref{tb:sp.eff} provides the efficiency comparison for the four split-plot designs under the four scenarios ($D$ efficiency for scenario (i) and GBD efficiency for the others).  
It shows that the two-level designs, $D_{sp1}$ and $D_{sp3}$, are more efficient when square terms are not potential important (scenarios (i) and (iii)). 
Design $D_{sp2}$ has lowest efficiency for scenarios (i), (ii), and (iii) but its efficiency is significantly higher than those of other three designs when the potential terms are chosen to be the four square terms (scenario (ii)). 

\begin{table}
\begin{center}
\caption{Variances of the estimated coefficients for the square terms in 15 models.}\label{tb:sp.var}

\begin{tabular}{cc|cccc}											
\hline											
Model	&	Square terms	&	$D_{sp1}$	&	$D_{sp2}$	&	$D_{sp3}$	&	$D_{sp4}$	\\
\hline											
1	&	A$^2$	&	-	&	2	&	-	&	2	\\
2	&	B$^2$	&	-	&	.5	&	-	&	1.38	\\
3	&	C$^2$	&	-	&	.5	&	-	&	1.38	\\
4	&	D$^2$	&	-	&	.5	&	-	&	1.38	\\
5	&	A$^2$,B$^2$	&	-	&	2,.5	&	-	&	2.17,1.5	\\
6	&	A$^2$,C$^2$	&	-	&	2,.5	&	-	&	2.17,1.5	\\
7	&	A$^2$,D$^2$	&	-	&	2,.5	&	-	&	2.17,1.5	\\
8	&	B$^2$,C$^2$	&	-	&	.5,.5	&	-	&	-	\\
9	&	B$^2$,D$^2$	&	-	&	.5,.5	&	-	&	-	\\
10	&	C$^2$,D$^2$	&	-	&	.5,.5	&	-	&	-	\\
11	&	A$^2$,B$^2$,C$^2$	&	-	&	2,.5,.5	&	-	&	-	\\
12	&	A$^2$,B$^2$,D$^2$	&	-	&	2,.5,.5	&	-	&	-	\\
13	&	A$^2$,C$^2$,D$^2$	&	-	&	2,.5,.5	&	-	&	-	\\
14	&	B$^2$,C$^2$,D$^2$	&	-	&	2,.5,.5	&	-	&	-	\\
15	&	A$^2$,B$^2$,C$^2$,D$^2$	&	-	&	2,.5,.5,.5	&	-	&	-	\\
\hline											
\end{tabular}											
\end{center}
\end{table}

In general, the GBD-optimal designs possess better projection properties. 
For instance, in scenario (ii), consider the 15 projective submodels including all primary terms and some square terms given in the second column of Table~\ref{tb:sp.var}. 
The variances of the estimated coefficients for the square terms in the models are given on the righthand side of Table~\ref{tb:sp.var}. 
For $D_{sp1}$ and $D_{sp3}$, all of the models are inestimable when square terms are included. 
That is because the two designs are two-level designs and, hence, the square terms in the models are completely aliasing with the intercept. 
For $D_{sp4}$, only seven submodels are estimable and the variances of the estimated coefficients for the square terms are larger than those for $D_{sp2}$. 
The GBD-optimal design, $D_{sp2}$, for scenario (ii) allows all of the 15 submodels to be estimated and has smallest variances for the estimated coefficients of the square terms.

We conduct a sensitivity study by setting $\eta_1$ from .1 to 10 to examine whether giving different values of $\eta_1$ will produce different resulting designs. 
Consequently, the algorithm reports the same optimal design (up to row, column, and level permutations) for each scenario even when $\eta_1\neq 1$. 
This is because we choose $\tau=10$, which is greater than $3\sigma_y$ for any $\eta_1$ less than or equal to 10 (see the discussion in Section~\ref{se:tau}).  
This study indicates that GBD-optimal split-plot designs are not sensitive to $\eta_1$ if the value of $\tau$ is chosen large enough.


\begin{table}
\begin{center}
\caption{Row-column table for the strip-plot design $D_{GBD-st}$.}\label{tb:strip1}																			
\begin{tabular}{rrrrrrrrrrrr}																								
\multicolumn{2}{c}{First stage factors}&&\multicolumn{9}{c}{Second stage factors}\\																								
\cline{1-2}\cline{4-12}																								
	$x_1^R$	&	$x_2^R$	&		&		&		&		&		&		&		&		&		&		\\
\hline																								
		&		&		&	$x_1^C$	&	-1	&	1	&	-1	&	1	&	-1	&	1	&	1	&	-1	\\
		&		&		&	$x_2^C$	&	1	&	-1	&	-1	&	1	&	1	&	-1	&	1	&	-1	\\
		&		&		&	$x_3^C$	&	-1	&	-1	&	-1	&	-1	&	1	&	1	&	1	&	1	\\
		&		&		&	$x_4^C$	&	-1	&	1	&	-1	&	1	&	1	&	-1	&	-1	&	1	\\
		&		&		&	$x_5^C$	&	1	&	-1	&	-1	&	1	&	-1	&	1	&	-1	&	1	\\
	-1	&	-1	&		&		&	$\surd$	&	$\surd$	&	$\surd$	&	$\surd$	&	$\surd$	&	$\surd$	&		&		\\
	1	&	-1	&		&		&	$\surd$	&	$\surd$	&	$\surd$	&	$\surd$	&		&		&	$\surd$	&	$\surd$	\\
	1	&	1	&		&		&	$\surd$	&	$\surd$	&		&		&	$\surd$	&	$\surd$	&	$\surd$	&	$\surd$	\\
	-1	&	1	&		&		&		&		&	$\surd$	&	$\surd$	&	$\surd$	&	$\surd$	&	$\surd$	&	$\surd$	\\
\hline																								
\end{tabular}																																														
\end{center}		
\end{table}

\begin{table}
\begin{center}
\caption{The GBD-optimal and $D$-optimal designs in Section~\ref{se:st}.}\label{tb:st.design}
\begin{tabular}{ccc}													
\begin{tabular}{rr|rrrrr}													
\multicolumn{7}{c}{(a) $D_{GBD-st}$}\\													
\hline													
$x_1^R$	&	$x_2^R$	&	$x_1^C$	&	$x_2^C$	&	$x_3^C$	&	$x_4^C$	&	$x_5^C$	\\
\hline													
-1	&	-1	&	-1	&	1	&	-1	&	-1	&	1	\\
-1	&	-1	&	1	&	-1	&	-1	&	1	&	-1	\\
-1	&	-1	&	-1	&	-1	&	-1	&	-1	&	-1	\\
-1	&	-1	&	1	&	1	&	-1	&	1	&	1	\\
-1	&	-1	&	-1	&	1	&	1	&	1	&	-1	\\
-1	&	-1	&	1	&	-1	&	1	&	-1	&	1	\\
1	&	-1	&	-1	&	1	&	-1	&	-1	&	1	\\
1	&	-1	&	1	&	-1	&	-1	&	1	&	-1	\\
1	&	-1	&	-1	&	-1	&	-1	&	-1	&	-1	\\
1	&	-1	&	1	&	1	&	-1	&	1	&	1	\\
1	&	-1	&	1	&	1	&	1	&	-1	&	-1	\\
1	&	-1	&	-1	&	-1	&	1	&	1	&	1	\\
1	&	1	&	-1	&	1	&	-1	&	-1	&	1	\\
1	&	1	&	1	&	-1	&	-1	&	1	&	-1	\\
1	&	1	&	-1	&	1	&	1	&	1	&	-1	\\
1	&	1	&	1	&	-1	&	1	&	-1	&	1	\\
1	&	1	&	1	&	1	&	1	&	-1	&	-1	\\
1	&	1	&	-1	&	-1	&	1	&	1	&	1	\\
-1	&	1	&	-1	&	-1	&	-1	&	-1	&	-1	\\
-1	&	1	&	1	&	1	&	-1	&	1	&	1	\\
-1	&	1	&	-1	&	1	&	1	&	1	&	-1	\\
-1	&	1	&	1	&	-1	&	1	&	-1	&	1	\\
-1	&	1	&	1	&	1	&	1	&	-1	&	-1	\\
-1	&	1	&	-1	&	-1	&	1	&	1	&	1	\\
\hline													
\end{tabular}													
&&													
\begin{tabular}{rr|rrrrr}													
\multicolumn{7}{c}{(b) $D_{AGJ-II}$}\\													
\hline													
$x_1^R$	&	$x_2^R$	&	$x_1^C$	&	$x_2^C$	&	$x_3^C$	&	$x_4^C$	&	$x_5^C$	\\
\hline													
-1	&	1	&	-1	&	-1	&	1	&	-1	&	1	\\
-1	&	1	&	1	&	1	&	1	&	1	&	-1	\\
-1	&	1	&	1	&	-1	&	1	&	1	&	1	\\
-1	&	1	&	-1	&	1	&	1	&	-1	&	-1	\\
-1	&	1	&	1	&	1	&	-1	&	-1	&	1	\\
-1	&	1	&	-1	&	-1	&	-1	&	1	&	-1	\\
1	&	-1	&	-1	&	-1	&	1	&	-1	&	1	\\
1	&	-1	&	1	&	1	&	1	&	1	&	-1	\\
1	&	-1	&	1	&	-1	&	1	&	1	&	1	\\
1	&	-1	&	-1	&	1	&	1	&	-1	&	-1	\\
1	&	-1	&	-1	&	1	&	-1	&	1	&	1	\\
1	&	-1	&	1	&	-1	&	-1	&	-1	&	-1	\\
-1	&	-1	&	-1	&	-1	&	1	&	-1	&	1	\\
-1	&	-1	&	1	&	1	&	1	&	1	&	-1	\\
-1	&	-1	&	1	&	1	&	-1	&	-1	&	1	\\
-1	&	-1	&	-1	&	-1	&	-1	&	1	&	-1	\\
-1	&	-1	&	-1	&	1	&	-1	&	1	&	1	\\
-1	&	-1	&	1	&	-1	&	-1	&	-1	&	-1	\\
1	&	1	&	1	&	-1	&	1	&	1	&	1	\\
1	&	1	&	-1	&	1	&	1	&	-1	&	-1	\\
1	&	1	&	1	&	1	&	-1	&	-1	&	1	\\
1	&	1	&	-1	&	-1	&	-1	&	1	&	-1	\\
1	&	1	&	-1	&	1	&	-1	&	1	&	1	\\
1	&	1	&	1	&	-1	&	-1	&	-1	&	-1	\\
\hline													
\end{tabular}													
\end{tabular}															
\end{center}
\end{table}

\subsection{Strip-plot designs}\label{se:st}
Strip-plot designs are used when experiments consist of two distinct process stages,  where the groups of semi-products produced in the first stage are applied to the second stage (see Miller, 1997, Federer and King, 2007, Vivacqua and Bisgaard, 2009, and  Arnouts et al., 2010).   
Such designs have three-stratum structures and can be expressed by row-column tables. 
Table~\ref{tb:strip1} is an example of the row-column table for a 24-run strip-plot design with two row factors $x_1^R$ and $x_2^R$ in the first stage and five column factors $x_1^C,\cdots,x_5^C$ in the second stage, where the check marks in the table present the structure and indicate the 24 runs of the strip-plot design. 
We present this design in Table~\ref{tb:st.design}(a) and denote it as $D_{GBD-st}$. 
This design is a GBD-optimal strip-plot design generated with $\tau=14$ and $\eta_l=1$ for $l=1,2,3$ when the intercept and main effects are the primary terms ($p=8$) and two-factors interactions are the potential terms ($q=21$).  
With the same strip-plot structure, Arnouts et al. (2010) provided a $D$-optimal design for the first-order model without consideration of the potential terms. The row-column table of the $D$-optimal strip-plot design is given in Table~II of Arnouts et al. (2010). We denote this design as $D_{AGJ-II}$ and present it in Table~\ref{tb:st.design}(b). 
It is impossible to generate a $D$-optimal design for the model including both main effects and two-factor interactions because the rank of the model matrix $\bX$ is smaller than the total number of the coefficients. 

\begin{figure}
 \begin{center}
 \includegraphics[width=5in]{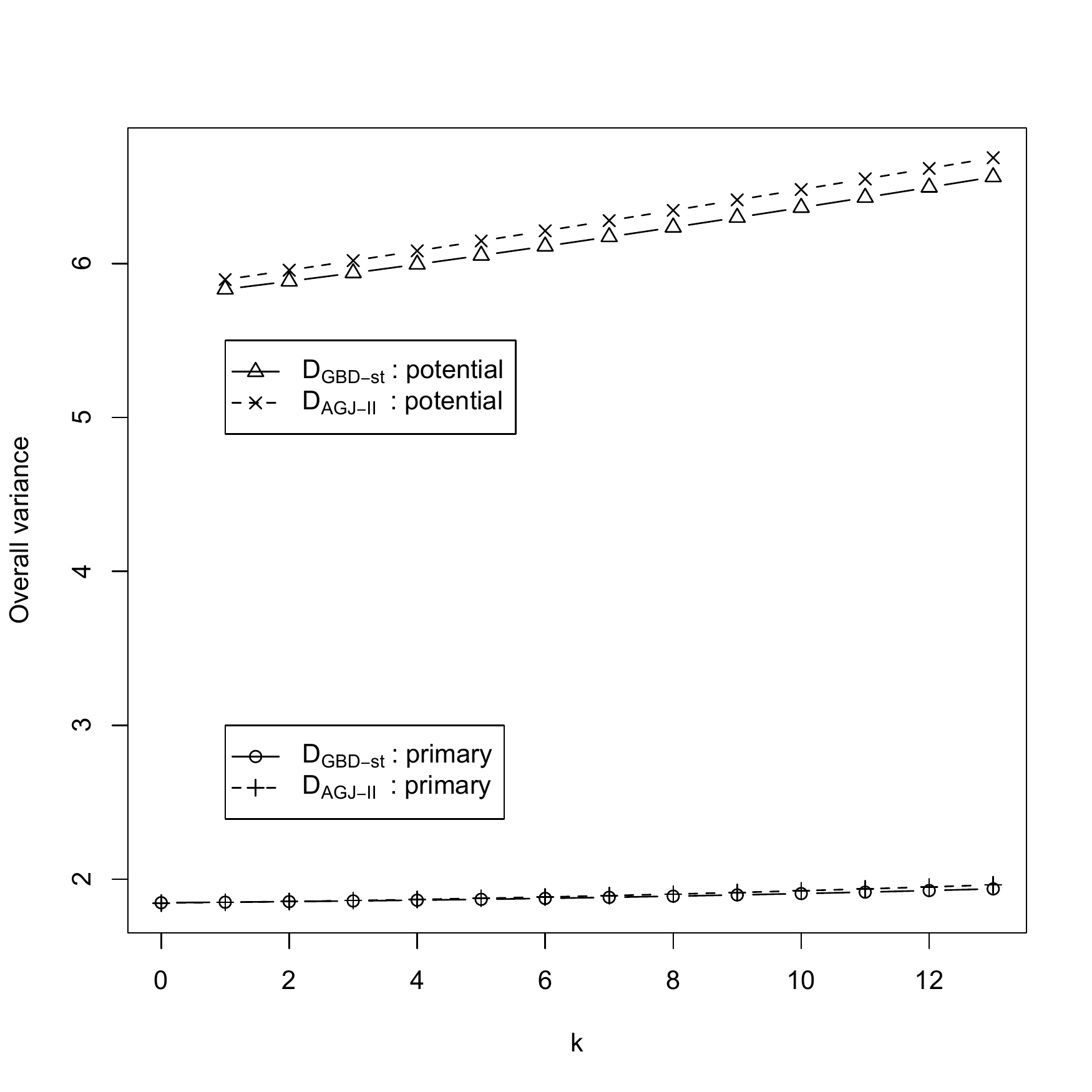}
 \end{center}
 \caption{Overall variance of the estimated coefficients for $D_{GBD-st}$ and $D_{AGJ-II}$.}\label{fi:st.var}
\end{figure}

Compare to the $D$-optimal design, $D_{AGJ-II}$, the GBD-optimal design, $D_{GBD-st}$, provides better performance on detecting active two-factor interactions. 
To observe this, we conduct the following study. 
For seven factors, there are total ${7\choose 2}=21$ two-factor interaction terms and, hence, there are total ${21\choose k}$ possible models including all main effects and $k$ two-factor interaction terms. Note that, not all of the ${21\choose k}$ models are estimable. 
For a given $k$, we calculate the variances of the estimated coefficients for all estimable models and take an average for each coefficient. 
In Figure~\ref{fi:st.var}, we present the overall variance (sum of the average variances) of the estimated coefficients of the primary and potential terms for $D_{GBD-st}$ and $D_{AGJ-II}$ when $0\leq k\leq13$.  
For the primary terms, it shows (on the bottom) that the overall variance of $D_{GBD-st}$ is slightly (not significantly) larger than that of $D_{AGJ-II}$ when $k\leq 1$ but it becomes smaller when $k\geq 2$. 
For the potential terms, it shows (on the top) that $D_{GBD-st}$ always has smaller overall variance than $D_{AGJ-II}$.  
This result indicates that the GBD-optimal design allows the precise estimation for all the primary terms while providing better detectability for the potential terms than the $D$-optimal design.

  
Through a sensitivity study by setting $\eta_1$ and $\eta_2$ varying from .1 to 10, we find that $D_{GBD-st}$ remains optimal even when $\eta_1$ and $\eta_2$ differ from 1.  
This result indicates that the GBD strip-plot design is not sensitive to $\eta_1$ and $\eta_2$ when $\tau>3\sigma_y$. 

\begin{table}
\begin{center}
\caption{A staggered-level structure and three GBD-optimal staggered-level designs for $\tau/\sigma_y=.0001$, $1$, and $3$.}\label{tb:sl1}
\begin{tabular}{cccc}										
\begin{tabular}{c|c|c}										
\multicolumn{3}{c}{(a) Structure}\\										
\hline										
WP$_{I}$		&	WP$_{II}$	&	SP	\\				
\hline										
\multirow{4}{*}{$\bw_1\otimes\bone_{4}$}&\multirow{2}{*}{$\bs_1\otimes\bone_{2}$}&$\bt_1$\\										
&&$\bt_2$\\										
\cline{2-2}										
&\multirow{4}{*}{$\bs_2\otimes\bone_{4}$}&$\bt_3$\\										
&&$\bt_4$\\										
\cline{1-1}										
\multirow{4}{*}{$\bw_2\otimes\bone_{4}$}&&$\bt_5$\\										
&&$\bt_6$\\										
\cline{2-2}										
&\multirow{4}{*}{$\bs_3\otimes\bone_{4}$}&$\bt_7$\\										
&&$\bt_8$\\										
\cline{1-1}										
\multirow{4}{*}{$\bw_3\otimes\bone_{4}$}&&$\bt_9$\\										
&&$\bt_{10}$\\										
\cline{2-2}										
&\multirow{4}{*}{$\bs_4\otimes\bone_{4}$}&$\bt_{11}$\\										
&&$\bt_{12}$\\										
\cline{1-1}										
\multirow{4}{*}{$\bw_4\otimes\bone_{4}$}&&$\bt_{13}$\\										
&&$\bt_{14}$\\										
\cline{2-2}										
&\multirow{4}{*}{$\bs_5\otimes\bone_{4}$}&$\bt_{15}$\\										
&&$\bt_{16}$\\										
\cline{1-1}										
\multirow{4}{*}{$\bw_5\otimes\bone_{4}$}&&$\bt_{17}$\\										
&&$\bt_{18}$\\										
\cline{2-2}										
&\multirow{2}{*}{$\bs_6\otimes\bone_{2}$}&$\bt_{19}$\\										
&&$\bt_{20}$\\										
\hline										
\end{tabular}										
&										
\begin{tabular}{r|r|rrr}										
\multicolumn{5}{c}{(b) $D_{sl1}$}\\										
\hline										
$w$		&	$s$	&	$t_1$	&	$t_2$	&	$t_3$	\\
\hline										
	1	&	1	&	1	&	-1	&	-1	\\
	1	&	1	&	-1	&	-1	&	1	\\
\cline{2-2}										
	1	&	-1	&	-1	&	-1	&	1	\\
	1	&	-1	&	1	&	1	&	1	\\
\cline{1-1}										
	-1	&	-1	&	-1	&	-1	&	-1	\\
	-1	&	-1	&	1	&	1	&	-1	\\
\cline{2-2}										
	-1	&	1	&	1	&	1	&	1	\\
	-1	&	1	&	1	&	-1	&	-1	\\
\cline{1-1}										
	1	&	1	&	1	&	1	&	-1	\\
	1	&	1	&	1	&	-1	&	1	\\
\cline{2-2}										
	1	&	1	&	-1	&	1	&	1	\\
	1	&	1	&	-1	&	-1	&	-1	\\
\cline{1-1}										
	-1	&	1	&	-1	&	-1	&	1	\\
	-1	&	1	&	-1	&	1	&	-1	\\
\cline{2-2}										
	-1	&	-1	&	-1	&	1	&	1	\\
	-1	&	-1	&	1	&	-1	&	1	\\
\cline{1-1}										
	1	&	-1	&	-1	&	1	&	-1	\\
	1	&	-1	&	1	&	-1	&	-1	\\
\cline{2-2}										
	1	&	1	&	1	&	1	&	1	\\
	1	&	1	&	-1	&	1	&	-1	\\
\hline										
\end{tabular}										
&										
\begin{tabular}{r|r|rrr}										
\multicolumn{5}{c}{(c) $D_{sl2}$}\\										
\hline										
$w$		&	$s$	&	$t_1$	&	$t_2$	&	$t_3$	\\
\hline										
	-1	&	-1	&	1	&	0	&	1	\\
	-1	&	-1	&	0	&	-1	&	0	\\
\cline{2-2}										
	-1	&	1	&	-1	&	1	&	1	\\
	-1	&	1	&	1	&	1	&	-1	\\
\cline{1-1}										
	1	&	1	&	-1	&	-1	&	1	\\
	1	&	1	&	1	&	1	&	1	\\
\cline{2-2}										
	1	&	-1	&	1	&	-1	&	1	\\
	1	&	-1	&	1	&	1	&	-1	\\
\cline{1-1}										
	-1	&	-1	&	-1	&	0	&	0	\\
	-1	&	-1	&	0	&	1	&	1	\\
\cline{2-2}										
	-1	&	1	&	1	&	-1	&	1	\\
	-1	&	1	&	-1	&	-1	&	-1	\\
\cline{1-1}										
	1	&	1	&	-1	&	1	&	-1	\\
	1	&	1	&	1	&	-1	&	-1	\\
\cline{2-2}										
	1	&	-1	&	-1	&	1	&	1	\\
	1	&	-1	&	-1	&	-1	&	-1	\\
\cline{1-1}										
	-1	&	-1	&	1	&	1	&	0	\\
	-1	&	-1	&	-1	&	-1	&	1	\\
\cline{2-2}										
	-1	&	-1	&	-1	&	1	&	-1	\\
	-1	&	-1	&	1	&	-1	&	-1	\\
\hline										
\end{tabular}										
&										
\begin{tabular}{r|r|rrr}										
\multicolumn{5}{c}{(d) $D_{sl3}$}\\										
\hline										
$w$		&	$s$	&	$t_1$	&	$t_2$	&	$t_3$	\\
\hline										
	0	&	0	&	-1	&	0	&	0	\\
	0	&	0	&	0	&	1	&	1	\\
\cline{2-2}										
	0	&	-1	&	1	&	1	&	-1	\\
	0	&	-1	&	-1	&	-1	&	-1	\\
\cline{1-1}										
	1	&	-1	&	-1	&	-1	&	1	\\
	1	&	-1	&	1	&	1	&	1	\\
\cline{2-2}										
	1	&	1	&	-1	&	1	&	1	\\
	1	&	1	&	-1	&	-1	&	-1	\\
\cline{1-1}										
	-1	&	1	&	-1	&	1	&	-1	\\
	-1	&	1	&	1	&	0	&	1	\\
\cline{2-2}										
	-1	&	-1	&	-1	&	1	&	1	\\
	-1	&	-1	&	1	&	-1	&	1	\\
\cline{1-1}										
	1	&	-1	&	-1	&	1	&	-1	\\
	1	&	-1	&	1	&	-1	&	-1	\\
\cline{2-2}										
	1	&	1	&	1	&	1	&	-1	\\
	1	&	1	&	1	&	-1	&	1	\\
\cline{1-1}										
	-1	&	1	&	1	&	-1	&	-1	\\
	-1	&	1	&	-1	&	-1	&	1	\\
\cline{2-2}										
	-1	&	0	&	0	&	0	&	-1	\\
	-1	&	0	&	1	&	1	&	0	\\
\hline										
\end{tabular}										
\end{tabular}																										
\end{center}
\end{table}

\subsection{Staggered-level designs}\label{se:sl}
Staggered-level designs are often used for experiments containing two classes of hard-to-change factors and the two classes of factors are reset at different points in time for cost-saving reasons. 
The two classes of factors are called the class-I and class-II whole-plot factors and their level combinations are call the class-I and class-II whole plots, respectively. 
%
A basic structure for staggered-level designs with five class-I whole plots of size 4 is given in Table~\ref{tb:sl1}(a), where $\otimes$ is the kronecker product, $\bone_f$ is an $f\times1$ vector of ones, and $\bw_i$, $\bs_j$, and $\bt_k$ are the $i$th class-I whole plot, the $j$th class-II whole plot, and the $k$th subplot, respectively. 
Typically, the first and last class-II whole plots are identical and have half size of the class-I whole plot.    
It has a three-stratum structure. 
More discussions about staggered-level designs can be found in Webb et al. (2004) and Arnouts and Goos (2012, 2015).

Table~9 of Arnouts and Goos (2015) listed a $D$-optimal staggered-level design with 20 runs, one class-I whole-plot factor, $w$, one class-II whole-plot factor, $s$, and two subplot factors, $t_1$ and $t_2$. 
This design is a three-level design selected by the $D$ criterion and can estimate all the coefficients for the second-order model constituted by the four factors.     
Assume that the experimenters now want to investigate one more subplot factor, $t_3$, without increasing the run size for maintaining experimental cost.  
Since the number of coefficients for the second-order model constituted by the five factors is larger than the rank of the model matrix, the $D$ criterion fails to determine an optimal design for this experiment.    
To overcome the singularity problem, we apply the GBD criterion with the primary terms including the intercept, main effects, and two-factor interactions, and the potential terms including square terms. 
To further investigate how the choice of $\tau$ affects the resulting designs, we generate three staggered-level designs $D_{sl1}$, $D_{sl2}$, and $D_{sl3}$ (listed in Table~\ref{tb:sl1}) by choosing $\tau=.0001\sigma_y$, $\sigma_y$, and $3\sigma_y$, respectively, where $\sigma_y=\sum_{l=1}^3\eta_l$ and $\eta_l=1$ for $l=1,\cdots,3$.  
Results show that when $\tau=.0001\sigma_y$, the GBD-optimal design, $D_{sl1}$, reduces to a two-level design. This $\tau$ is close to zero, which implies that effects of the square terms are very small and can be ignored. Hence, $D_{sl1}$ is equivalent to the $D$-optimal design for the model including the primary terms only. 
When $\tau=\sigma_y$, the GBD-optimal design, $D_{sl2}$, has levels at 0 for the factors in the lower stratum. 
This design allows some square terms to be detected but not all.  
When $\tau=3\sigma_y$, which implies that effects of the square terms are likely to be significant, the GBD-optimal design, $D_{sl3}$, has levels at 0 for all factors and, hence, allows the square of each factor to be detected.     

\begin{figure}
 \begin{center}
 \includegraphics[width=6in]{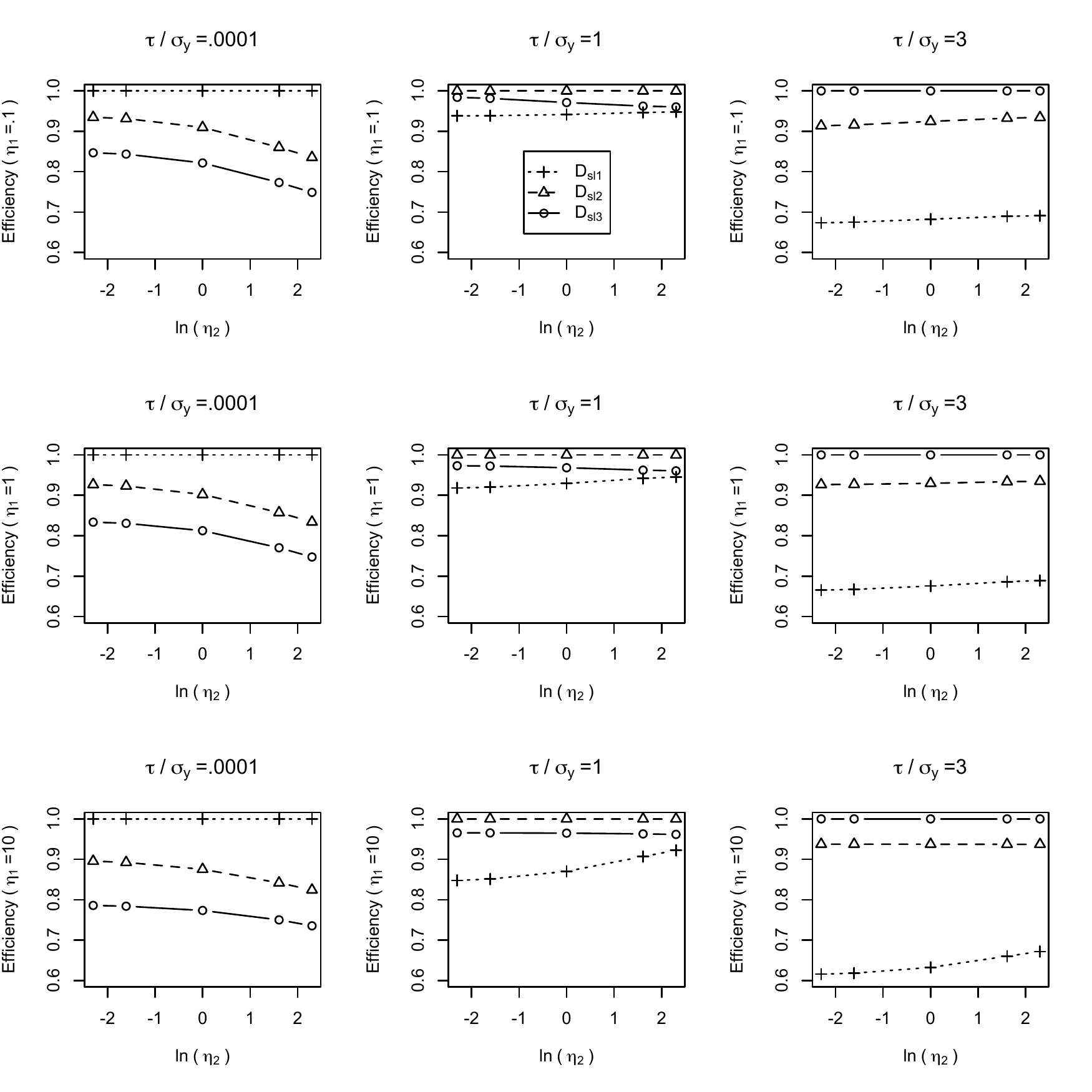}
 \end{center}
 \caption{Efficiency comparison for $D_{sl1}$, $D_{sl2}$, and $D_{sl3}$ when $\eta_1$ and $\eta_2$ varying from .1 to 10 and $\tau/\sigma_y=.0001$, 1, and 3.}\label{fi:sl.eff}
\end{figure}

We consider different settings for $\eta_1$ and $\eta_2$ 
and compute the GBD efficiency for $D_{sl1}$, $D_{sl2}$, and $D_{sl3}$. 
Results for $\tau/\sigma_y=.0001$, $1$, and $3$ are presented in the first, second, and third columns of Figure~\ref{fi:sl.eff}, respectively. 
It shows that, although different values of $\eta_1$ and $\eta_2$ are given, designs $D_{sl1}$, $D_{sl2}$, and $D_{sl3}$ remain optimal (with highest efficiency) for $\tau/\sigma_y=.0001$, $1$, and $3$, respectively.   
This result indicates that the GBD-optimal staggered-level designs are not sensitive to $\eta_l$'s when $\tau$ and $\sigma_y$ keep the same ratio.

\section{Concluding remarks}\label{se:cr}
In this study, we extend the result of the Bayesian approach proposed by DuMouchel and Jones (1994).  
Under the framework of multistratum structures, we develop the GBD (generalized Bayesian $D$) criterion, which generalizes the original Bayesian $D$ criterion in DuMouchel and Jones (1994) and can be applied to select single-stratum and multistratum designs for various experiments. 
Since the GBD criterion considers primary and potential terms simultaneously, the GBD-optimal design can reduce dependence on an assumed model. 
By adding prior information to the effects of the primary and potential terms, the GBD criterion circumvents the singularity problem when the rank of the model matrix is not large enough to estimate all the coefficients. 
GBD-optimal designs in general provide smaller variances for the estimated coefficients of the potential terms. 
They allow the precise estimation for all the primary terms while providing better detectability for the potential terms.   
%
With the GBD criterion, many existing works developed based on the original Bayesian $D$ criterion can be further extended. 
For instance, with the GBD criterion, the Bayesian two-stage designs for mixture models proposed by Lin et al. (2000) can be extended and applied to multistratum experiments.  
These extensions are valuable and worth to study and will be considered in our future research. 

\end{document}